\documentclass[aps,prl,twocolumn,showpacs,superscriptaddress,floatfix]{revtex4}

\usepackage{graphicx}
\usepackage{amssymb}

\begin{document}

\title{Muon spin rotation and relaxation in the superconducting ferromagnet UCoGe}

\author{A. de Visser}
\email{devisser@science.uva.nl} \affiliation{Van der Waals -
Zeeman Institute, University of Amsterdam, Valckenierstraat~65,
1018 XE Amsterdam, The Netherlands}
\author{N. T. Huy}
\altaffiliation{Current address: Hanoi Advanced School of Science
and Technology, Hanoi University of Technology, 1 Dai Co Viet,
Hanoi, Vietnam} \affiliation{Van der Waals - Zeeman Institute,
University of Amsterdam, Valckenierstraat~65, 1018 XE Amsterdam,
The Netherlands}
\author{A. Gasparini}
\affiliation{Van der Waals - Zeeman Institute, University of
Amsterdam, Valckenierstraat~65, 1018 XE Amsterdam, The
Netherlands}
\author{D. E. de Nijs}
\affiliation{Van der Waals - Zeeman Institute, University of
Amsterdam, Valckenierstraat~65, 1018 XE Amsterdam, The
Netherlands}
\author{D. Andreica}
\altaffiliation{Current address: Faculty of Physics, Babes-Bolyai
University, 400084 Cluj-Napoca, Romania} \affiliation{Laboratory
for Muon-Spin Spectroscopy, Paul Scherrer Institute, CH-5232
Villigen, Switzerland}
\author{C. Baines}
\affiliation{Laboratory for Muon-Spin Spectroscopy, Paul Scherrer
Institute, CH-5232 Villigen, Switzerland}
\author{A. Amato}
\affiliation{Laboratory for Muon-Spin Spectroscopy, Paul Scherrer
Institute, CH-5232 Villigen, Switzerland}

\date{\today}

\begin{abstract}
We report zero-field muon spin rotation and relaxation
measurements on the superconducting ferromagnet UCoGe. Weak
itinerant ferromagnetic order is detected by a spontaneous muon
spin precession frequency below the Curie temperature $T_C = 3$~K.
The $\mu^+$ precession frequency persists below the bulk
superconducting transition temperature $T_{sc} = 0.5$~K, where it
measures a local magnetic field $B_{loc} = 0.015$~T. The amplitude
of the $\mu$SR signal provides unambiguous proof for
ferromagnetism present in the whole sample volume. We conclude
ferromagnetism coexists with superconductivity on the microscopic
scale.

\end{abstract}


\maketitle

Recently, it was shown~\cite{Huy-PRL-2007} that UCoGe belongs to a
new family of ferromagnetic metals which become superconducting at
low temperatures. One of the most intriguing properties of these
superconducting ferromagnets (SCFMs) is that both ordering
phenomena coexist. Such a coexistence is incompatible with the
standard BCS theory for superconductivity (SC), because the
ferromagnetic (FM) exchange field prevents phonon-mediated
formation of singlet Cooper pairs~\cite{Berk-PRL-1966}. However,
for weak itinerant ferromagnets, like UCoGe, an alternative route
is offered by spin fluctuation
models~\cite{Fay-PRB-1980,Lonzarich-CUP-1997}, in which critical
magnetic fluctuations near a magnetic phase transition can mediate
unconventional SC by pairing electrons in a spin triplet state.
Superconductivity not mediated by
phonons~\cite{Monthoux-Nature-2007} attracts a wide interest as it
plays a central role in the quest to unravel high temperature SC
in the cuprates~\cite{Bednorz-ZPhysB-1986} and the
oxypnictides~\cite{Kamihara-JAmChemSoc-2008} discovered recently.
The SCFMs discovered so far are UGe$_2$ (under
pressure)~\cite{Saxena-Nature-2000}, UIr (under
pressure)~\cite{Akazawa-JPCM-2004}, URhGe~\cite{Aoki-Nature-2001}
and UCoGe. The coexistence of itinerant FM and SC in these
compounds marks a clear distinction with other families of
superconductors, like Chevrel phases~\cite{Fischer-ApplPhys-1978}
and borocarbides~\cite{Cava-Nature-1994}, in which local-moment
(anti)ferromagnetism and SC exist, but expel each other. In the
case of UGe$_2$~\cite{Pfleiderer-PRL-2002} and
URhGe~\cite{Levy-Science-2005} experimental evidence is at hand
that SC is driven by critical magnetic fluctuations near a
magnetic transition between two polarized phases. In contrast,
UCoGe may present the first example of SC stimulated by critical
fluctuations associated with a FM quantum critical point.

FM order in UCoGe with a Curie temperature $T_{C} = 3$~K was first
observed by magnetization $M(T,B)$ measurements on polycrystalline
samples~\cite{Huy-PRL-2007}. The small (polycrystalline averaged)
ordered moment $m_{0}$= 0.03 $\mu_B$ and the low coercive field of
$\sim 1$~mT indicate magnetism is weak. The analysis of the
magnetization by means of Arrott plots shows UCoGe is an itinerant
FM. This is further corroborated by specific heat data, which show
the entropy associated with the magnetic transition is small (0.3
\% of $R$ln2). Subsequent $M(T,B)$ data taken on single crystals
reveal a strong uniaxial magnetic anisotropy: the moment $m_{0} =
0.07 ~\mu_B$ points along the orthorhombic $c$
axis~\cite{Huy-PRL-2008}. In the FM state SC occurs with a
resistive transition temperature $T_{sc} = 0.8$~K for the best
polycrystalline samples. Measurements of the upper critical field
$B_{c2}$ provide evidence for unconventional SC, characterized by
an equal-spin pairing triplet state and a SC gap function with
point nodes along the direction of the uniaxial
moment~\cite{Huy-PRL-2008}.

Although FM order is a robust property of our poly and
single-crystalline samples, it has not been probed directly via
magnetization below 1.8 K~\cite{Huy-PRL-2007}. Moreover, the
$M(T,B)$ data alone cannot exclude that the small moment is due to
a reduced volume part of the sample exhibiting FM order with a
larger magnetic moment. Evidence obtained so far for the
coexistence of SC and FM stems mainly from macroscopic
measurements, notably thermal expansion~\cite{Huy-PRL-2007}. The
sizes of the steps in the coefficient of linear thermal expansion
at $T_C$ and $T_{sc}$ indicate magnetism and SC occur in the whole
sample, however, again, reduced volume fractions cannot be ruled
out. Therefore, it is of paramount importance to investigate the
magnetic response of UCoGe on a microscopic scale. The $\mu$SR
technique is extremely suitable for this
purpose~\cite{Amato-RMP-1997}. Positive muons $\mu^+$, when
implanted in the sample, act as a sensitive probe of the local
field, which permits to discern magnetically inequivalent sample
regions. In this Letter we present a $\mu$SR study on UCoGe which
provides unambiguous proof for magnetic order in the whole sample
volume. FM coexists with SC below $T_{sc}$. Our work is the first
$\mu$SR study conducted on a SC itinerant FM below its SC
transition temperature. The muon response in the SC state provides
evidence for a spontaneous vortex phase in zero field.
Furthermore, our results provide insight in the intricate problem
of how the superconductor accommodates the FM structure.

A batch of polycrystalline UCoGe was prepared as described in
Ref.~\cite{Huy-PRL-2007}. Measurements of the electrical
resistivity, $R(T)$, show the residual resistance ratio
$RRR=R(300$~K$)/R(1$~K$) \approx 30$ and the FM and SC transitions
occur at 3.0 K and 0.8 K (onset), respectively (see Fig.~3a).
Zero-field (ZF) $\mu$SR experiments were performed at the Paul
Scherrer Institute in the General Purpose Spectrometer (GPS) using
a $^4$He flow cryostat in the temperature range $T = 1.6-10$~K and
in the Low Temperature Facility (LTF) using a dilution
refrigerator for $T = 0.02-5.5$~K. Two identical $\mu$SR samples
were prepared. Each sample consisted of four thin slices
(thickness $0.8$~mm, area $6$x$10$~mm$^2$) that were cut from the
annealed button by spark erosion. The surface layer, defected by
spark erosion, was removed by polishing. The total sample area was
$12$x$20$~mm$^2$ and matched the cross-sectional area of the muon
beam in order to eliminate contributions from the sample holder.
One sample was measured in the GPS and the other in the LTF. In
addition we carried out ZF $\mu$SR experiments on polycrystalline
URh$_{0.4}$Co$_{0.6}$Ge, which has a Curie temperature $T_C =
20.0$~K~\cite{Sakarya-JALCOM-2008} and a (polycrystalline
averaged) ordered moment $m_{0} =
0.22~\mu_B$~\cite{Huy-PhDThesis-2008}.

ZF $\mu$SR is a well known technique to study magnetically ordered
phases. Polarized muons are implanted into a sample where their
spins ${\bf S}_{\mu}$ ($S_{\mu} = 1/2$) precess in the local
magnetic field ${\bf B}_{loc}$ until they decay. ${\bf
S}_{\mu}(t)$ is monitored through the decay positron. By measuring
the asymmetric distribution of emitted positrons the time
evolution of the polarization ${\bf P}_{\mu}(t)$ can be deduced.
$P_{\mu}(t)$ is defined as the projection of ${\bf P}_{\mu}(t)$
along the direction of initial polarization $P_{\mu}(t) = {\bf
P}_{\mu}(t) \cdot {\bf P}_{\mu}(0) / P_{\mu}(0) = G(t)
P_{\mu}(0)$, where $G(t)$ reflects the normalized muon-spin
autocorrelation function $G(t)= \langle {\bf S}_{\mu}(t) \cdot
{\bf S}_{\mu}(0)  \rangle / {S_{\mu}(0)}^2$. The muon
depolarization function $G(t)$ extracted from the $\mu$SR spectra
contains the information on the averaged value, distribution and
time evolution of the internal magnetic fields and is to be
compared with selected theory functions.
\begin{figure}
\includegraphics[width=6.5cm]{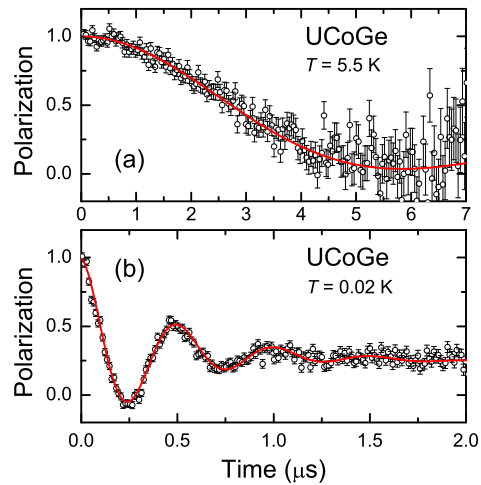}
%Musr spectra of polycrystalline UCoGe
%at 5.5 K and 0.02 K.
\caption{Typical zero-field $\mu$SR spectra in polycrystalline
UCoGe: (a) in the paramagnetic phase at $T= 5.5$~K (solid line fit
to Eq.~1), and (b) in the ferromagnetic phase at $T = 0.02$~K,
well below the superconducting transition temperature (solid line
fit to Eq.~2). }
\end{figure}

A typical ZF $\mu$SR spectrum on UCoGe in the paramagnetic state
at $T=5.5$~K is shown in Fig.~1a. This spectrum is best described
by the standard Kubo-Toyabe function:
\begin{equation}
G_{KT} (t)= \frac{1}{3} + \frac{2}{3} (1- \Delta_{KT} ^{2} t^{2})
\exp (- \frac{1}{2} \Delta_{KT} ^{2} t^{2} )
\end{equation}

The Kubo-Toyabe function describes the muon depolarization due to
an isotropic Gaussian distribution of static internal fields
centered at zero field. $\Delta_{KT} = \gamma_{\mu} \sqrt{\langle
B^2 \rangle}$ is the Kubo-Toyabe relaxation rate, with
$\gamma_{\mu}$ the muon gyromagnetic ratio ($\gamma_{\mu} /2 \pi =
135.5~$MHz/T) and $\langle B^2 \rangle$ the second moment of the
field distribution. Spectra taken in the paramagnetic phase show
$\Delta_{KT}$ is $0.30 \pm 0.01~\mu$s$^{-1}$ and is independent of
temperature. This indicates the depolarization is not due to
electronic magnetic moments, but rather due to static nuclear
moments of the $^{59}$Co atoms ($\mu_{Co} = 5.23~\mu_N$, the spin
$I = 7/2$ and the abundance is $100~ \% $).

In the FM state, a clear spontaneous $\mu^+$ precession frequency
$\nu$ is observed. A typical spectrum, taken at $T=0.02$~K is
shown in Fig.~1b. This response has an electronic origin and is
best described by the depolarization function of an isotropic
polycrystalline magnet with a Lorentzian field distribution:
\begin{equation}
G_{M} (t)= \frac{1}{3} \exp (-\lambda_1 t) + \frac{2}{3} \exp
(-\lambda_2 t) \cos (2 \pi \nu t + \phi )
\end{equation}

The first term corresponds to an average of 1/3 of the muons that
sense a local field parallel to their initial polarization and
therefore do not precess, where $\lambda_1$ is a measure for the
internal spin dynamics perpendicular to the muon spin. The `2/3
term' probes the local field which results in a precession of the
muon spin with frequency $\nu$ ($\phi$ is a phase factor) and
$\lambda_2$ is a measure for both static and dynamical spin
effects. By fitting the spectrum taken at $T=0.02$~K (Fig.~1b) to
Eq.~2 we obtain the spontaneous frequency $\nu = 1.972 \pm
0.004$~MHz, and the damping rates $\lambda_1 = 0.152 \pm
0.006~\mu$s$^{-1}$ and $\lambda_2 = 2.372 \pm 0.004~\mu$s$^{-1}$.
The measured value of $\lambda_1$ confirms the `1/3' term is due
to the sample and muons do not stop in the sample holder. As will
be demonstrated below, the spontaneous $\mu^+$ precession
frequency characterizes the weak itinerant FM order. The observed
frequency corresponds to a local field $B_{loc}$ $\sim 0.015$~T at
the muon localization site.

In order to investigate the evolution of the magnetic volume
fraction in the temperature regime of the FM transition we have
fitted the $\mu$SR spectra to a two-component depolarization
function:
\begin{equation}
G (t) = A_{magn} G_{M} (t) + A_{para} G_{KT} (t)
\end{equation}

Here $A_{magn}$ and $A_{para}$ represent the volume fractions of
the ferro- and paramagnetic phases, respectively, with
normalization $A_{magn} + A_{para} = 1$. In Fig.~2 we show the
deduced temperature variation $A_{magn} (T)$ and $A_{para} (T)$.
The magnetic transition is relatively broad in our polycrystalline
samples. Below $\sim 1.5$~K $A_{magn} (T) \rightarrow 1$ and
magnetism is present in the whole sample volume. The ZF asymmetry
is within error bar equal to the asymmetry measured in transverse
field (TF = 50 G) at temperatures of 5.5 K (LTF) and 10 K (GPS),
which confirms the full signal amplitude is involved.

\begin{figure}
\includegraphics[width=6.5cm]{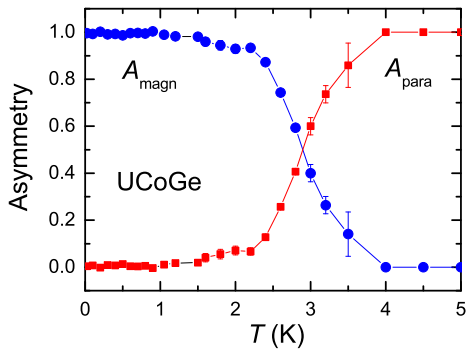}
%Normalized asymmetry of polycrystalline UCoGe
\caption{Temperature variation of the normalized asymmetries of
the ferromagnetic and paramagnetic phase in UCoGe. Below $\sim
1.5$~K magnetism is observed in the whole sample volume. The solid
lines are to guide the eye.}
\end{figure}

In Fig.~3 we show $\nu (T)$, which tracks the magnetization
$M(T)$~\cite{Huy-PRL-2007}. The $\nu (T)$ data can be well fitted
to the phenomenological order-parameter function $\nu (T) = \nu_0
(1-(T/T^{*})^{\alpha})^{\beta}$. The spontaneous frequency
vanishes near $T^* =3.02$~K $\approx T_C $ and $\nu_0 = 1.98$~MHz.
Notice the critical exponent $\alpha = 2.3$ deviates from the
standard value 3/2 for spin waves, while $\beta = 0.4$ is larger
than the theoretical value 5/16 for a 3D Ising-like
FM~\cite{Essam-JChemPhys-1963}.

\begin{figure}
\includegraphics[width=6.5cm]{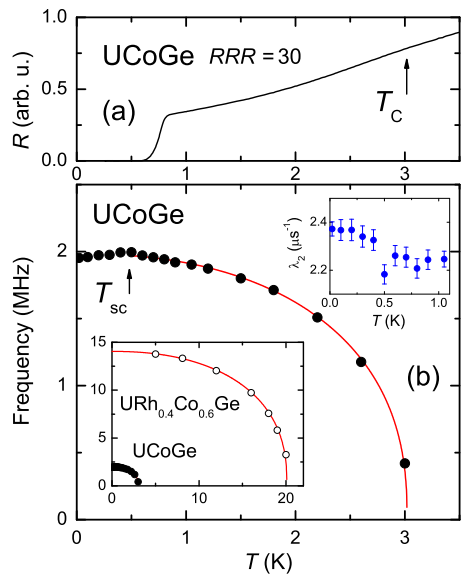}
%Temperature variation of oscillation frequency of UCoGe
\caption{(a) Resistivity versus temperature of polycrystalline
UCoGe. (b) Spontaneous muon precession frequency $\nu (T)$ of
UCoGe and (lower inset) URh$_{0.4}$Co$_{0.6}$Ge. The solid lines
represent fits to a phenomenological order parameter function (see
text). Notice the 2\% decrease of $\nu (T)$ below the bulk
$T_{sc}$ of $0.5$~K~\cite{Huy-PRL-2007}. Upper inset: $\lambda
_{2}(T)$ of UCoGe.}
\end{figure}

In order to further explore the $\mu$SR spectra in the magnetic
state we have carried out ZF muon spin rotation experiments on
UCoGe doped with Rh. Upon substituting Rh for Co the Curie
temperature and the ordered moment steadily increase and reach the
values $T_C = 20.0$~K~\cite{Sakarya-JALCOM-2008} and $m_{0} =
0.22~\mu_B$~\cite{Huy-PhDThesis-2008} for 40 at.~\% Rh. ZF $\mu$SR
spectra on polycrystalline URh$_{0.4}$Co$_{0.6}$Ge were collected
in the GPS for $T= 5 - 25$~K. In the FM phase the muon
depolarization is described by Eq.~2, but best fits were obtained
with a Gaussian damping factor $\exp (- \frac{1}{2} \sigma_G ^2
t^2)$ for the `2/3 term'. From the amplitude of the signal we
conclude magnetism is observed in the whole sample volume. The
main feature of the spectra is the spontaneous $\mu ^+$ precession
frequency below $T_C$, which attains a value of 14.1~MHz at the
lowest temperature (see inset of Fig.~3). $\nu (T)$ follows an
order-parameter variation with $T^* = 20.1$~K $\approx T_C$. The
values $\alpha = 2.2$ and $\beta = 0.4$ are similar to the ones
derived from the magnetization~\cite{Huy-PhDThesis-2008}, and
identical, within the error bars, to the values obtained for
UCoGe. The precession frequency $\nu (0)$ is a factor $\sim 7$
larger than the one of UCoGe and scales with the ratio of the
ordered moments ($\sim 7$) deduced from the
dc-magnetization~\cite{Huy-PhDThesis-2008}. We conclude the
analysis of the ZF $\mu$SR spectra of URh$_{0.4}$Co$_{0.6}$Ge
corroborates the results for UCoGe.

Next we discuss the possible muon localization site. UCoGe
crystallizes in the orthorhombic TiNiSi type
structure~\cite{Lloret-PhDThesis-1988,Canepa-JALCOM-1996} (space
group $P_{nma}$), where the U, Co and Ge atoms occupy the 4c sites
(Wyckoff notation). The most probable localization sites are the
high symmetry interstitial 4a and 4b positions. By calculating
$\Delta_{KT}$ due to the $^{59}$Co nuclear moments ($T
> T_C$) one can in principle determine the stopping
site~\cite{Schenck-AHBook-1985}. From a search along high symmetry
directions in the crystal leads we tentatively assign the $\mu ^+$
localization site to the 4a position, where $\Delta_{KT} = 0.34~
\mu $s$^{-1}$~\cite{stoppingsite}, close to the experimental value
of $0.30 \pm 0.01~ \mu $s$^{-1}$. The local field in the FM phase
(in ZF) is given by $B_{loc} = B_{dip}+B_{c}+B_{L}$, where
$B_{dip}$ is the dipolar sum of the ordered moments, $B_{c}$ is
the contact hyperfine field and $B_{L}=\mu _{0} M_{s}/3$ the
Lorentz field with $M_s$ the saturation magnetization (see for
example Ref.~\cite{Schenck-HandbookMM-1995}). For a precise
determination of the different terms Knight shift experiments on a
single crystal are required. However, dipolar calculations yield
the following estimates: with $m_{0} = 0.07 ~\mu_B ~\|~ c$ at the
U atoms~\cite{Huy-PRL-2008}, we derive for the 4a site $B_{dip} =
0.015$~T, $B_{L}= 0.005$~T, and by comparing to the experimental
value $B_{loc}$ of 0.015 T, a negative contact field $B_{c} =
-0.005$~T.

We now have a detailed understanding of the ZF muon spin rotation
data and proceed to make a number of important conclusions. First
of all, the data provide unambiguous proof for ferromagnetism
present in the whole sample volume. The spontaneous $\mu^+$
precession signal persists with the same asymmetry below the SC
transition temperature (see Fig.~3). Thus FM and SC coexist.
Interestingly, in the SC state the precession frequency shows a
small decrease of about 2~\%. This decrease is measured by the
whole muon ensemble. It provides a first indication of the
interplay of magnetism and SC in UCoGe. The decrease of $\nu$
below $T_{sc}$ is possibly due to induced screening currents that
lead to a small reduction of the local field. On the other hand,
the damping factor $\lambda_2$ slightly increases when passing
through $T_{sc}$ (see Fig.3). Such a broadening of the field
distribution is expected when a spontaneous vortex
phase~\cite{Tachiki-SSC-1980} is formed. This indicates the local
field is larger than the lower critical field $B_{c1}$ and the
sample is always in the mixed state. Our results also shed light
on the intricate question of how the SC state accommodates FM
order. The strong uniaxial magnetic anisotropy~\cite{Huy-PRL-2008}
excludes the formation of a modulated or spiral magnetic structure
when passing through $T_{sc}$, like observed in
ErRh$_4$B$_4$~\cite{Sinha-PRL-1982}. This is corroborated by the
minor changes in the $\mu$SR spectra below $T_{sc}$. It has been
proposed that the SC transition should be accompanied by a
redistribution of FM domains
~\cite{Anderson-PR-1959,Sonin-PRB-2002}, however, such a
redistribution cannot be detected in our polycrystal data. $\mu$SR
experiments on single-crystals are needed to elucidate these
issues further.

We emphasize the FM state is observed in zero applied magnetic
field. Recent magnetization measurements on single crystals ($RRR
= 20$) led to the claim that a magnetic field of the order of a
few mT is needed to stabilize FM
order~\cite{PoltierovaVejprova-arXiv-2008}. This is at variance
with our results~\cite{Huy-PRL-2008} and indicates metallurgy of
UCoGe is an important issue. This might also be of relevance for
understanding recent Co NMR and NQR data obtained on powdered
samples~\cite{Ohta-JPSJ-2008}. The analysis reveals a
two-component nuclear spin relaxation rate $1/T_1$, yielding
evidence for an inhomogeneous sample and a SC volume fraction of
30~\%.

In summary, we have used the $\mu$SR technique to investigate the
weak itinerant FM phase in UCoGe. Below $T_C = 3$~K a spontaneous
$\mu^+$ precession frequency appears, which probes a local field
$B_{loc}$ of 0.015 T when $T \rightarrow 0$. Muons most likely
localize at the 4a site, as deduced from calculations of the
Kubo-Toyabe relaxation rate. The $\mu^+$ precession frequency
persists below $T_{sc}$, where FM and SC coexist on the
microscopic scale.

This work was part of the research program of the Dutch Foundation
FOM. Part of this work was carried out at the Swiss Muon Source
S$\mu$S (PSI, Villigen, Switzerland) with support of the European
Commission under the 6th Framework Programme through the COST
Action P16 ECOM and the Key Action Strengthening the European
Research Area, Research Infrastructures (contract
RII3-CT-2003-505925).


\begin{thebibliography}{99}
\bibitem{stoppingsite} For the 4b site $\Delta_{KT} = 0.20~ \mu $s$^{-1}$. The second moment
calculation includes quadrupolar coupling effects. We assume a
radial electric field gradient and neglect the small contribution
from nuclear moments at Ge atoms.
\end{thebibliography}
\end{document}